\documentclass{article}
\usepackage{spconf,amsmath,graphicx}
\usepackage{amssymb}
\usepackage{multirow} 
\usepackage{CJKutf8} 
\usepackage{url} 

\usepackage{caption}
\usepackage{subcaption}
\usepackage{flushend}

\usepackage{subcaption}
\usepackage{booktabs}

\title{MERTECH: INSTRUMENT PLAYING TECHNIQUE DETECTION using SELF-SUPERVISED PRETRAINED MODEL with MULTI-TASK FINETUNING}
\name{
    \begin{tabular}{c}
    Dichucheng Li $^{1*}$\thanks{$^*$The authors contributed equally to this work. 
    Wei Li, Emmanouil Benetos, and Fan Xia are corresponding authors.
    This work was supported by NSFC(62171138). 
    Yinghao Ma is a research student at the UKRI Centre for Doctoral Training in Artificial Intelligence and Music, supported by UK Research and Innovation [grant number EP/S022694/1]. Emmanouil Benetos is supported by a RAEng/Leverhulme Trust Research Fellowship [grant number LTRF2223-19-106].}, 
    Yinghao Ma \textsuperscript{2,\includegraphics[scale=0.03]{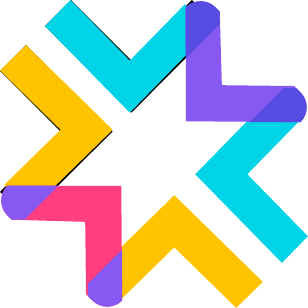}*}, 
    Weixing Wei $^1$, Qiuqiang Kong $^{3}$, Yulun Wu $^1$,\\
    Mingjin Che $^4$, Fan Xia $^{4\dag}$, Emmanouil Benetos $^{2\dag}$, and Wei Li $^{1,5\dag}$
    \end{tabular}
}
\address{
    $^1$ School of Computer Science and Technology, Fudan University, Shanghai, China \\
    $^2$ Centre for Digital Music, Queen Mary University of London, UK\\
    \includegraphics[scale=0.03]{map-logo-c.pdf} Multimodal Art Projection Research Community\\
    $^3$ The Chinese University of Hong Kong (CUHK)\\
    $^4$ College of Experimental Art, Sichuan Conservatory of Music, Sichuan, China\\
    $^5$ Shanghai Key Laboratory of Intelligent Information Processing, Fudan University, China\\
 }

\begin{document}
\begin{CJK*}{UTF8}{gbsn}
	\ninept
	\maketitle
	\begin{abstract}

        Instrument playing techniques (IPTs) constitute a pivotal component of musical expression. However, the development of automatic IPT detection methods suffers from limited labeled data and inherent class imbalance issues.
        In this paper, we propose to apply a self-supervised learning model pre-trained on large-scale unlabeled music data and finetune it on IPT detection tasks. This approach addresses data scarcity and class imbalance challenges.
        Recognizing the significance of pitch in capturing the nuances of IPTs and the importance of onset in locating IPT events, we investigate multi-task finetuning with pitch and onset detection as auxiliary tasks.
        Additionally, we apply a post-processing approach for event-level prediction, where an IPT activation initiates an event only if the onset output confirms an onset in that frame. Our method outperforms prior approaches in both frame-level and event-level metrics across multiple IPT benchmark datasets. Further experiments demonstrate the efficacy of multi-task finetuning on each IPT class. 
        \footnote{Code: https://github.com/LiDCC/MERTech}
        
	\end{abstract}

    \begin{figure*}[t]
    \centering
    \includegraphics[width=17.8cm]{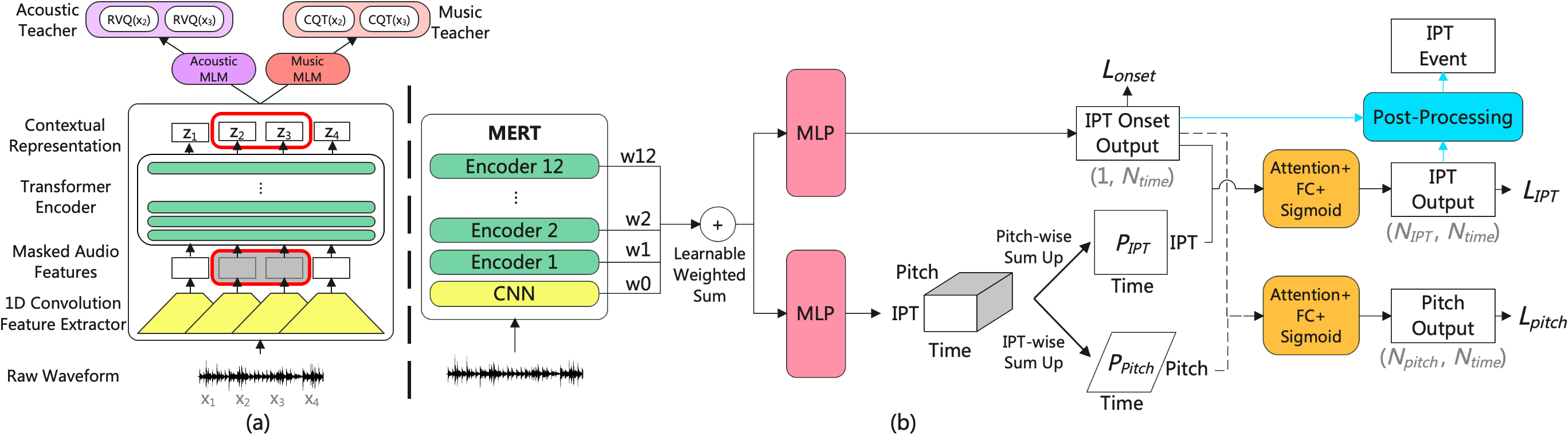}
    \caption{The overall architecture of \texttt{MERTech}. (a) Stage I: Pre-training MERT-v1-95M. (b) Stage II: Multi-task finetuning for instrument playing technique (IPT) detection, pitch detection, and IPT onset detection. The blue part of (b) denotes Stage III: Post-processing. Characters in parentheses indicate the shapes of the outputs. ``$N_ {IPT}$", ``$N_ {pitch}$", ``$N_ {time}$" are the number of IPT classes, pitch classes, and output frames.}
    \label{fig:model}
    \end{figure*}
    
    \section{Introduction}
    \label{sec:intro}

    Instrument playing techniques (IPTs), such as vibratos and glissandos, play a crucial role in enhancing the expressiveness of musical performances. The goal of IPT detection is to classify IPT types and identify their positions within an audio signal. The modeling and detection of IPTs benefit other tasks in music information retrieval (MIR), such as instrument classification \cite{lostanlen2018extended}, performance analysis \cite{yang2017computational}, and automatic music transcription (AMT) \cite{TENT, huang2023note}.
    
    Early works on IPT detection mostly used machine learning methods, combined with hand-crafted features. Wang et al. \cite{wang2022adaptive} employed support vector machines (SVMs) on adaptive scattering features. Chen et al. \cite{chen2015electric} utilized SVMs on a combination of  Mel-frequency cepstral coefficients (MFCCs), pitch contour features, and timbre features. Advancements in deep learning have led to the increasing utilization of deep neural networks \cite{foldedcqt, li2022playing}. Su et al. \cite{TENT} extended the work of \cite{chen2015electric} by substituting SVMs with convolutional neural networks (CNNs). Huang et al. \cite{huang2023note} introduced U-net models for joint prediction of notes, IPTs, note states, and IPT groups. Li et al. \cite{li2023frame} proposed a hybrid model of multi-scale convolution and self-attention mechanism. 
    
    However, deep learning approaches often demand extensive datasets comprising high-quality labeled audio tracks. The main challenge in the IPT detection task lies in the scarcity of large-scale datasets. While there exist relatively large datasets for clip-level IPT classification \cite{lostanlen2018extended}, our paper focuses on frame-level and event-level IPT detection in real-world instrumental performance audio sequences. These datasets \cite{huang2023note, wang2022adaptive, li2023frame} typically only include around 2 hours of audio each due to the labor-intensive and expert-level manual labeling process. 
    Besides, IPTs serve as embellishments in music performance, resulting in the majority of notes being labeled as `normal' without any specific IPT assignments. Consequently, the inherent class imbalance issue is present within the IPT detection task. 
    In situations of limited data and class imbalance, supervised learning is susceptible to problems like overfitting, poor performance on minority class samples, and restricted generalization.

    To address data scarcity and class imbalance problems in IPT detection, we propose employing self-supervised learning (SSL) models. SSL models are pre-trained on large-scale unlabeled corpora, and the knowledge gained from pre-training can be transferred to downstream tasks either by using the model as a feature extractor or by finetuning the entire model. Previous research has demonstrated the effectiveness of SSL models in low-resource scenarios \cite{hsu2021hubert} and their robustness to data imbalance \cite{liu2021self}. While initially introduced in natural language processing (NLP), an increasing number of SSL models have recently emerged in the field of acoustic music, such as MULE \cite{mccallum2022supervised}, MapMusic2Vec \cite{li2022map}, and MERT \cite{li2023mert}. Among these models, MERT excels in sequence labeling tasks where other models encounter difficulties due to the lack of frame-level representations or are too cumbersome to train \cite{yuan2023marble}. Additionally, MERT has been shown to be a particularly promising model for transfer learning \cite{yamamoto2023toward} and has achieved the best results in performance-level tasks (like vocal technique recognition) \cite{yuan2023marble}. Hence, we opt to finetune MERT for the IPT detection task.

    To further improve IPT detection performance, we explore multi-task finetuning for IPT detection, where the auxiliary task is pitch detection and IPT onset detection. SSL offers robust feature embeddings that can enhance multi-task finetuning for downstream tasks \cite{chen2021speech}. Pitch plays a crucial role in capturing IPT nuances while predicting IPT onsets aids in locating IPT events.

    The main contributions of this paper are as follows: 1) We propose finetuning a pre-trained SSL model on IPT detection, demonstrating its effectiveness, generalizability, and capability to alleviate class imbalance on three benchmark IPT datasets of Guzheng, guitar, and Chinese bamboo flute (CBF). 2) We further explore multi-task finetuning for IPT detection with pitch detection and IPT onset detection as auxiliary tasks, proving its effectiveness on two IPT datasets with labeled pitch information. 3) We propose a post-processing approach, where an IPT activation initiates an event only if the onset output confirms an onset in that frame, resulting in significant improvement in event-level metrics.

    \section{Method}
    The overall framework of our proposed model, \texttt{MERTech}, is illustrated in Fig. \ref{fig:model}. We first review the structure and the pre-training strategy of MERT. Then we elaborate on the proposed multi-task finetuning method and post-processing strategy.
    \subsection{Structure and Pre-training Process of MERT-v1-95M}
    We employ MERT-v1-95M \cite{li2023mert}, an acoustic music pre-trained model comprising a CNN-based feature extractor coupled with a transformer-based contextual network as shown in Fig. \ref{fig:model}(a). When handling the initial 24 kHz audio input, the CNN transforms it into a 75 Hz feature representation. This representation is then processed by a 12-layer Transformer, resulting in a 75 Hz contextual hidden variable with 768 dimensions.
    
    The pre-training approach of MERT-v1 involves segment-wise masking of the feature representation. The model is then tasked with reconstructing two music features using guidance from a musical teacher and an acoustic teacher, following the well-established masked language model (MLM) paradigm. The musical teacher imparts pitch-related knowledge by Constant-Q Transform (CQT), while the acoustic teacher imparts knowledge related to acoustic features, each of which includes 8 embeddings modeled by Residual Vector Quantization (RVQ), generated from EnCodec \cite{defossez2022high}.
    
    
    The pre-training of MERT-v1 is executed on a dataset comprising roughly 160k hours of unlabeled music data, predominantly of Western origin. In this paper, we employed the base (95M) size models, which are trained with a 1K hours subset.
    
    \subsection{Multi-task Finetuning on IPT Detection, Pitch Detection, and IPT Onset Detection}
    We present the structure of the downstream models in Fig. \ref{fig:model}(b). Our downstream model is composed of two branches: the output of the top branch is the IPT onset output ($\hat{Y}_{onset}$), while the output of the bottom branch is the raw posteriorgrams of pitch ($P_{pitch}$) and IPT ($P_{IPT}$). Subsequently, two prediction refinement sub-networks, each employing a self-attention layer and a fully connected (FC) layer, combine $\hat{Y}_{onset}$ with $P_{pitch}$ and $P_{IPT}$, respectively, resulting in frame-level pitch output ($\hat{Y}_{pitch}$) and IPT output ($\hat{Y}_{IPT}$).
    
    The input waveform is trimmed to 5 seconds and resampled to 24k Hz, aligning with the preprocessing step during the pre-training stage. Subsequently, we utilize a weighted sum of CNN and 12 Transformer encoder layer outputs from MERT as the input for the downstream model. The effectiveness of the weighted sum approach lies in its utilization of information from various semantic levels across different layers \cite{chen2023exploring}. We employ 13 learnable weight values, each assigned to the output of its respective layer.
    
    A one-layer 512-unit Multilayer Perceptron (MLP) is applied to each branch. A dropout layer with a rate of 0.2 and a ReLU layer are applied after the first linear layer of each MLP. The last fully-connected (FC) layer is time-distributed (applied to each frame).
    
    IPT onset detection differs from note onset detection. A single note may include multiple consecutive IPTs, and conversely, an IPT event may span multiple notes. For example, a guitar slide technique involves sliding a finger across frets to reach another note. Thus, IPT onset detection identifies the start of each IPT, not individual notes. The last FC layer in the MLP of the top branch has a target size of 1, predicting the presence of IPT onsets of each time frame.
    
    In the bottom branch, the target size of the last FC layer in the MLP is $N_{IPT}$ × $N_{pitch}$, where $N_{IPT}$ is the number of IPT classes, and $N_{pitch}$ is the number of pitch classes. After that, the output is reshaped to an order-3 tensor $D$ of size $N_{time}$ × $N_{IPT}$ × $N_{pitch}$, where $N_{time}$ indicates the number of output frames. Then $D$ is summed up along IPT axis and pitch axis to get $P_{pitch}$ and $P_{IPT}$, respectively. The multi-task approach of predicting $P_{pitch}$ and $P_{IPT}$ within a single branch, utilizing the same MLP, is inspired by previous work \cite{huang2022improving}. We prefer this method to a multi-head architecture commonly used in multi-task learning because it maintains a close alignment between pitch and IPT information throughout the process. For instance, a Guzheng glissando involves discrete pitches within pentatonic scales. Thus, jointly predicting pitch and IPTs is expected to yield superior results compared to separate predictions.
    
    After obtaining $\hat{Y}_{onset}$, $P_{pitch}$, and $P_{IPT}$, we detach $\hat{Y}_{onset}$ firstly to avoid backward propagation, then concatenate it with $P_{pitch}$ and $P_{IPT}$ along the frequency dimension, respectively, to predict $\hat{Y}_{pitch}$ and $\hat{Y}_{IPT}$. The attention mechanism employed in the refinement modules is self-attention. In terms of the Q-K-V convention used to describe a self-attention block \cite{vaswani2017attention}, all components-Q, K, and V-are derived from the input of the refinement module. 

    To address downstream tasks, we finetune the models in a supervised manner. Following the strategy in \cite{hsu2021hubert}, we freeze the parameters of the CNN-based feature extractor in MERT during finetuning while updating other parts of the model.

    The total loss function is a weighted sum of the IPT onset loss ($L_{onset}$), IPT loss ($L_{IPT}$), and pitch loss ($L_{pitch}$). Specifically, $L_{pitch}$ and $L_{onset}$ are computed using frame-level binary cross entropy (BCE) loss between predictions and ground truths. For calculating $L_{IPT}$, we follow \cite{li2023frame}, employing the weighted BCE loss as proposed in \cite{citeloss} to further address the class imbalance issue. The final loss is calculated as Eq.\ref{eq:loss}.

    \begin{equation}\label{eq:loss}
    L = \lambda_{1} L_{IPT} + \lambda_{2} L_{pitch} + \lambda_{3} L_{onset}
    \end{equation}
    where $\lambda_{1}$, $\lambda_{2}$, and $\lambda_{3}$ are adjustable parameters, set to 1.0, 0.5, and 0.5, respectively, determined via coarse hyperparameter search.
    
    \subsection{Post-Processing}
    The post-processing step is exclusively employed during testing to transform frame-level IPT output into event-level IPT prediction. We use a threshold of 0.5 to binarize the onset output. Similar to \cite{onsetsAndframes}, an activation from the IPT output is permitted to initiate an IPT event only if the onset output confirms the presence of an onset in that particular frame.
    
    The difference is, in piano transcription \cite{onsetsAndframes}, the onset output shares the same shape as the pitch output. Thus, when allowing a pitch event to be initiated by an activation from the pitch output, it is essential for the onset output not only to predict an onset in that frame but also for the corresponding pitch. Given that predicting IPT type at the start of a note is more challenging than predicting pitch, our focus is on predicting IPT onsets without specifying the IPT type in the onset output. This minimizes the occurrence of False Negative IPT events after post-processing.

    \section{Experiments}
    
    \subsection{Datasets}
    Three different datasets are selected to train and evaluate the proposed models in our experiments. 
    
    Guzheng\_Tech99 \cite{li2023frame} contains 99 polyphonic Guzheng solo pieces recorded by 2 professional Guzheng players, totaling 151.1 minutes. The dataset includes 7 classes of playing techniques, namely vibrato, point note, upward portamento, downward portamento, glissando, tremolo, and plucks.

    EG-Solo \cite{huang2023note} contains 76 electric guitar solos with polyphonic backing tracks from YouTube, totaling 40 minutes. The dataset includes 9 classes of playing techniques: normal, slide, bend, vibrato (aka trill), mute, pull (aka pull-off), harmonic, hammer (aka hammer-on), and tap.
 
    CBFdataset \cite{wang2022adaptive} comprises 80 monophonic CBF performances recorded by 10 professional performers, totaling 2.6 hours. The dataset includes 7 playing techniques: vibrato, tremolo, trill, flutter-tongue, acciaccatura, portamento, and glissando.
    
    \subsection{Experimental Setup}

    The experiments follow the default dataset splits in their original papers. Specifically, the CBFdataset is divided in a 8:2 ratio based on CBF players, and a 5-fold cross-validation is conducted. 
    
    We finetune the model using stochastic gradient descend (SGD) with momentum 0.9, an initial learning rate of 0.001, a batch size of 10, a gradient clipping L2-norm of 3, and a cosine learning rate scheduler. 
    \subsection{Metrics}
    Because the research on IPT detection is still in its early stages, most papers propose new datasets for experimentation, leading to a lack of standardized metrics. In this context, we report four typical metrics to comprehensively evaluate the performance of each model.
    
    We present frame-level and event-level F1-scores with a default 50ms onset tolerance using the mir\_eval library \cite{raffel2014mir}. In terms of averaging options for metrics calculation, we provide both micro-averaging and macro-averaging results \cite{mesaros2016metrics}. Micro-averaging assigns equal weight to individual frame decisions, maintaining consistency with music transcription metrics \cite{onsetsAndframes}. Macro-averaging computes the average class-wise performance, placing emphasis on the performance for smaller classes in the problem.

    \section{Results}
    In this section, we conduct ablation studies to showcase our design benefits. We then compare our methods with state-of-the-art (SOTA) approaches on three datasets. Finally, we create a histogram of F1-scores for each IPT in Guzheng\_Tech99 dataset to analyze the impact of multi-task learning and transfer learning on each IPT.

    Firstly, we conduct ablation studies to showcase our design benefits. We consider the following four variant models: (a) \texttt{IPT+Pitch+Onset} eliminates the post-processing step from \texttt{MERTech}, so \texttt{MERTech} can be seen as \texttt{IPT+Pitch+Onset+pp} (``pp" refers to post-processing). (b) \texttt{IPT+Pitch} retains only the bottom branch of \texttt{MERTech} for simultaneous pitch and IPT prediction. (c) \texttt{IPT\_finetune} is a single-task version of \texttt{MERTech}, using a one-layer 512-unit MLP in the downstream model for exclusive IPT prediction. (d) \texttt{IPT\_probing} shares the model structure of \texttt{IPT\_finetune}, but with frozen parameters in MERT and only updates the parameters in the downstream model.

    \begin{table}[htb]
 \begin{center}
 \resizebox{.75\width}{!}{
 \renewcommand{\arraystretch}{1.3}
 \begin{tabular}{c |c c c c}
 \toprule[2pt]
 \multirow{2}{*}{Model}  & \multicolumn{2}{c}{FRAME-LEVEL} & \multicolumn{2}{c}{EVENT-LEVEL} \\
 
 \cmidrule(r){2-3} \cmidrule(r){4-5}
 
  & MI-F1 (\%) &  MA-F1(\%) &  MI-F1 (\%) &  MA-F1(\%) \\
 
\midrule[1.5pt]
 & \multicolumn{4}{c}{Guzheng\_Tech99}\\

\hline
IPT\_probing  & 81.2 & 60.3 & 19.9 & 16.5  \\
 
\hline
IPT\_finetune & 91.0 & 81.8 & 49.0 & 54.7 \\

\hline
IPT+Pitch & \pmb{91.7} & 83.4 & 50.0 & 54.8 \\

\hline
IPT+Pitch+Onset & 91.4 & \pmb{84.5} & 50.2 & 57.0 \\

\hline
MERTech & 90.0 & 80.4 & \pmb{91.6} & \pmb{76.1} \\

\midrule[1.5pt]
 & \multicolumn{4}{c}{EG-Solo}\\

\hline
IPT\_probing   & 57.3 & 21.5 & 12.0 & 10.5  \\
 
\hline
IPT\_finetune & 64.3 & 29.9 & 27.6 & 17.0 \\

\hline
IPT+Pitch & 65.2 & \pmb{31.6} & 23.8 & 15.6\\

\hline
IPT+Pitch+Onset & \pmb{66.1} & 31.3 & 36.7 & 20.1 \\

\hline
MERTech  & 62.2 & 28.7 & \pmb{54.3} & \pmb{24.6} \\

\midrule[1.5pt]
 & \multicolumn{4}{c}{CBFdataset}\\

\hline
IPT\_probing   & 87.3 & 73.2 & 17.1 & 22.0  \\
 
\hline
IPT\_finetune & \pmb{92.8} & \pmb{83.5} & \pmb{61.5} & \pmb{60.3} \\

 \bottomrule[2pt]
 \end{tabular}
 }
\end{center}
 \caption{Ablation studies with Frame-level and event-level F1-scores using micro-averaging (MI-F1) and macro-averaging (MA-F1) on three datasets.}
 \label{tab:ablation}
\end{table}

\begin{table}[htb]
 \begin{center}
 \resizebox{.75\width}{!}{
 \renewcommand{\arraystretch}{1.3}
 \begin{tabular}{c |c c c c}
 \toprule[2pt]
 \multirow{2}{*}{Model}  & \multicolumn{2}{c}{FRAME-LEVEL} & \multicolumn{2}{c}{EVENT-LEVEL} \\
 
 \cmidrule(r){2-3} \cmidrule(r){4-5}
 
  & MI-F1 (\%) &  MA-F1(\%) &  MI-F1 (\%) &  MA-F1(\%) \\
 
\midrule[1.5pt]
 & \multicolumn{4}{c}{Guzheng\_Tech99}\\

\hline
GZFNO  \cite{li2022playing}  & 68.7 & - & - & - \\
 
\hline
Multi-scale \cite{li2023frame} & 86.5 & 69.3 & 38.8 & 38.9 \\

\hline
MERTech & \pmb{90.0} & \pmb{80.4} & \pmb{91.6} & \pmb{76.1}\\

\midrule[1.5pt]
 & \multicolumn{4}{c}{EG-Solo}\\

\hline
Solola \cite{TENT}  & 44.1 & - & - & -   \\

\hline
NATSolo \cite{huang2023note} & 57.9 & - & - & -  \\

\hline
MERTech  & \pmb{62.2} & \pmb{28.7} & \pmb{54.3} & \pmb{24.6} \\

\midrule[1.5pt]
 & \multicolumn{4}{c}{CBFdataset}\\

\hline
AST \cite{wang2022adaptive}  & - & 79.9 & - & - / 63.9$^{*}$   \\

\hline
IPT\_finetune & \pmb{92.8} & \pmb{83.5} & \pmb{61.5} & \pmb{60.3 / 72.7$^{*}$} \\

 \bottomrule[2pt]
 \end{tabular}
 }
\end{center}
 \caption{Comparison with state-of-the-art approaches on three datasets. “-” refers to “not reported by the previous papers”. Results with ``*" were measured with a 200ms onset tolerance for fair comparison with prior studies.}
 \label{tab:compare}
\end{table}

    As shown in Table \ref{tab:ablation}, when comparing \texttt{IPT+Pitch+Onset} with \texttt{MERTech}, not using post-processing yields a slight increase in frame-level metrics, while causing a significant 41.4\%, 19.1\%, 17.6\%, and 4.5\% decrease in event-level (which better aligns with human judgment \cite{ycart2020investigating}) micro-F1 and macro-F1 for Guzheng\_Tech99 and EG-Solo, respectively. Similar patterns emerge in other music transcription studies \cite{onsetsAndframes}. \texttt{IPT+Pitch+Onset} outperforms \texttt{IPT+Pitch} notably in event-level metrics, especially in the EG-Solo dataset, showing the importance of onset information for identifying the position of IPTs in instrument performance audio with background music interference. \texttt{IPT+Pitch} surpasses \texttt{IPT\_finetune} in most cases, evidencing the importance of pitch information. Lastly, \texttt{IPT\_finetune} outperforms \texttt{IPT\_probing} across all datasets reveals the effectiveness of our finetuning strategy, aligning the SSL model for downstream tasks.

    We compare the performance of our proposed methods with previous approaches, as shown in Table \ref{tab:compare}. We train and evaluate \texttt{MERTech} on the Guzheng\_Tech99 and EG-Solo datasets, which include labels for both IPT and pitch. However, due to the absence of pitch labels in the CBFdataset, we present the results of the \texttt{IPT\_finetune} model for this dataset. Our method significantly outperforms all previously published results, confirming the effectiveness of our model in IPT detection. Particularly our method improves 52.8\% and 37.2\% in event-level micro-F1 and macro-F1, respectively, in the Guzheng\_Tech99 dataset. Besides, our model excels in both frame-level and event-level macro-F1, showcasing its capability to alleviate class imbalance issues. Furthermore, by comparing \texttt{IPT\_finetune} with \texttt{AST} \cite{wang2022adaptive} in the CBFdataset, we show the generalizability of our model across a heterogeneous test set and training set (different performers). Lastly, it is vital to note the distributional variance between the Western-centric pre-trained dataset and downstream datasets, including two with Chinese traditional music. This underscores the strong potential of MERT for low-resource music data scenarios of underrepresented styles.
    
\begin{figure}[t]
    \centering
    \includegraphics[width=8.3cm]{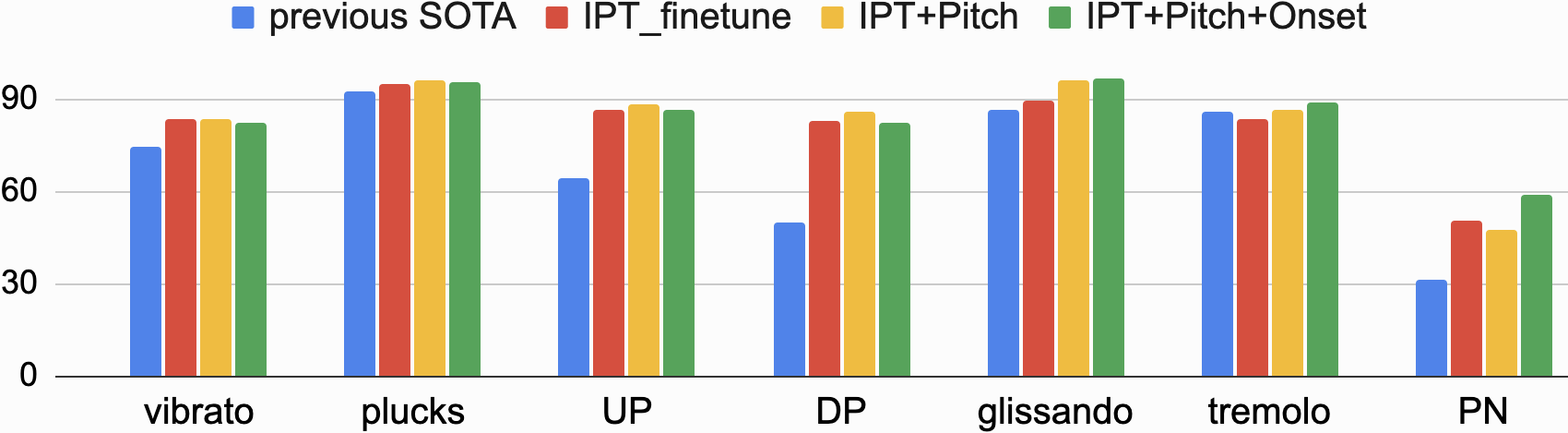}
    \caption{The frame-level F1-score for each IPT in Guzheng\_Tech99 dataset. ``UP" is ``Upward Portamento", ``DP" is ``Downward Portamento", and ``PN" is ``Point Note".}
    \label{fig:histogram}
    \end{figure}

    To analyze the multi-task learning and transfer learning impact on each IPT, we create a histogram of frame-level F1-scores for each IPT in Guzheng\_Tech99 dataset. We exclude event-level results due to the potential interference of onset threshold choice and \texttt{MERTech} outcomes due to rule-based post-processing interference. In Fig. \ref{fig:histogram}, \texttt{IPT\_finetune} shows more balanced performance across categories than previous SOTA \cite{li2023frame}, demonstrating its capability to address class imbalance issues. Besides, \texttt{IPT+Pitch} excels in upward/downward portamento, both involving pitch sliding. It also outperforms \texttt{IPT\_finetune} significantly in glissando, as Guzheng glissando relate to pitch, generating discrete pitches within pentatonic scales. \texttt{IPT+Pitch+Onset} yields the best results in glissando, tremolo, and point note (PN). The first two IPTs encompass multiple string-plucking onset sounds, while PN resembles a specific type of vibrato with a singular pitch change. Onset information aids in distinguishing PN from vibrato by identifying the IPT boundaries, and the similarity between PN and vibrato is why PN has the lowest F1-score.

    \section{Conclusion}
    In this paper, we propose to apply an SSL model pre-trained on extensive unlabeled music data and finetune it for IPT detection to address data scarcity and class imbalance challenges. We also explore multi-task finetuning with pitch and IPT onset detection as auxiliary tasks. Additionally, we introduce a post-processing approach for event-level prediction, where an IPT activation initiates an event only if the onset output confirms an onset in that frame. Our method outperforms existing approaches in frame-level and event-level metrics across three IPT benchmark datasets. Further experiments demonstrate the efficacy of multi-task finetuning on each IPT class. In the future, we aim to extend the application of music SSL model to other low-resource conditions and also explore other methods like semi-supervised learning to tackle low-resource challenges in this task.

    \section{Acknowledgments}
     We would like to thank Changhong Wang and Tung-Sheng Huang for their assistance in dataset provision and invaluable advice. 
	

 \bibliographystyle{IEEEbib}
	\bibliography{strings,refs}

\begin{thebibliography}{10}

\bibitem{lostanlen2018extended}
Vincent Lostanlen, Joakim And{\'e}n, and Mathieu Lagrange,
\newblock ``Extended playing techniques: the next milestone in musical instrument recognition,''
\newblock in {\em Proceedings of the International Conference on Digital Libraries for Musicology}, 2018.

\bibitem{yang2017computational}
Luwei Yang,
\newblock {\em Computational modelling and analysis of vibrato and portamento in expressive music performance},
\newblock Ph.D. thesis, Queen Mary University of London, 2017.

\bibitem{TENT}
Ting-Wei Su, Yuan-Ping Chen, Li~Su, and Yi-Hsuan Yang,
\newblock ``Tent: Technique-embedded note tracking for real-world guitar solo recordings,''
\newblock {\em Transactions of the International Society for Music Information Retrieval}, vol. 2, no. 1, pp. 15--28, 2019.

\bibitem{huang2023note}
Tung-Sheng Huang, Ping-Chung Yu, and Li~Su,
\newblock ``Note and playing technique transcription of electric guitar solos in real-world music performance,''
\newblock in {\em IEEE International Conference on Acoustics, Speech and Signal Processing, ICASSP}, 2023.

\bibitem{wang2022adaptive}
Changhong Wang, Emmanouil Benetos, Vincent Lostanlen, and Elaine Chew,
\newblock ``Adaptive scattering transforms for playing technique recognition,''
\newblock {\em IEEE/ACM Transactions on Audio, Speech, and Language Processing}, vol. 30, 2022.

\bibitem{chen2015electric}
Yuan-Ping Chen, Li~Su, and Yi-Hsuan Yang,
\newblock ``Electric guitar playing technique detection in real-world recording based on f0 sequence pattern recognition.,''
\newblock in {\em Proceedings of the 16th International Society for Music Information Retrieval Conference, ISMIR}, 2015, pp. 708--714.

\bibitem{foldedcqt}
Jean-Francois Ducher and Philippe Esling,
\newblock ``Folded cqt rcnn for real-time recognition of instrument playing techniques,''
\newblock in {\em Proceedings of the 20th International Society for Music Information Retrieval Conference, ISMIR}, 2019, pp. 708--714.

\bibitem{li2022playing}
Dichucheng Li, Yulun Wu, Qinyu Li, Jiahao Zhao, Yi~Yu, Fan Xia, and Wei Li,
\newblock ``Playing technique detection by fusing note onset information in guzheng performance,''
\newblock in {\em Proceedings of the 23rd International Society for Music Information Retrieval Conference, ISMIR, Bengaluru, India}, 2022, pp. 314--320.

\bibitem{li2023frame}
Dichucheng Li, Mingjin Che, Wenwu Meng, Yulun Wu, Yi~Yu, Fan Xia, and Wei Li,
\newblock ``Frame-level multi-label playing technique detection using multi-scale network and self-attention mechanism,''
\newblock in {\em IEEE International Conference on Acoustics, Speech and Signal Processing, ICASSP}, 2023, pp. 1--5.

\bibitem{hsu2021hubert}
Wei-Ning Hsu, Benjamin Bolte, Yao-Hung~Hubert Tsai, Kushal Lakhotia, Ruslan Salakhutdinov, and Abdelrahman Mohamed,
\newblock ``Hubert: Self-supervised speech representation learning by masked prediction of hidden units,''
\newblock {\em IEEE/ACM Transactions on Audio, Speech, and Language Processing}, vol. 29, pp. 3451--3460, 2021.

\bibitem{liu2021self}
Hong Liu, Jeff~Z HaoChen, Adrien Gaidon, and Tengyu Ma,
\newblock ``Self-supervised learning is more robust to dataset imbalance,''
\newblock in {\em NeurIPS 2021 Workshop on Distribution Shifts: Connecting Methods and Applications}, 2021.

\bibitem{mccallum2022supervised}
Matthew~C McCallum, Filip Korzeniowski, Sergio Oramas, Fabien Gouyon, and Andreas Ehmann,
\newblock ``Supervised and unsupervised learning of audio representations for music understanding,''
\newblock in {\em Proceedings of the 23th International Society for Music Information Retrieval Conference, ISMIR}, 2022.

\bibitem{li2022map}
Yizhi Li, Ruibin Yuan, Ge~Zhang, Yinghao MA, Chenghua Lin, Xingran Chen, Anton Ragni, Hanzhi Yin, Zhijie Hu, Haoyu He, et~al.,
\newblock ``Map-music2vec: A simple and effective baseline for self-supervised music audio representation learning,''
\newblock in {\em ISMIR 2022 Hybrid Conference}, 2022.

\bibitem{li2023mert}
Yizhi Li, Ruibin Yuan, Ge~Zhang, Yinghao Ma, Xingran Chen, Hanzhi Yin, Chenghua Lin, Anton Ragni, Emmanouil Benetos, Norbert Gyenge, et~al.,
\newblock ``Mert: Acoustic music understanding model with large-scale self-supervised training,''
\newblock {\em arXiv preprint arXiv:2306.00107}, 2023.

\bibitem{yuan2023marble}
Ruibin Yuan, Yinghao Ma, Yizhi Li, Ge~Zhang, Xingran Chen, Hanzhi Yin, Le~Zhuo, Yiqi Liu, Jiawen Huang, Zeyue Tian, et~al.,
\newblock ``Marble: Music audio representation benchmark for universal evaluation,''
\newblock {\em arXiv preprint arXiv:2306.10548}, 2023.

\bibitem{yamamoto2023toward}
Yuya Yamamoto,
\newblock ``Toward leveraging pre-trained self-supervised frontends for automatic singing voice understanding tasks: Three case studies,''
\newblock {\em arXiv preprint arXiv:2306.12714}, 2023.

\bibitem{chen2021speech}
Yi-Chen Chen, Shu-wen Yang, Cheng-Kuang Lee, Simon See, and Hung-yi Lee,
\newblock ``Speech representation learning through self-supervised pretraining and multi-task finetuning,''
\newblock in {\em AAAI 2022 Workshop on Self-supervised Learning for Audio and Speech Processing}, 2021.

\bibitem{defossez2022high}
Alexandre D{\'e}fossez, Jade Copet, Gabriel Synnaeve, and Yossi Adi,
\newblock ``High fidelity neural audio compression,''
\newblock {\em arXiv preprint arXiv:2210.13438}, 2022.

\bibitem{chen2023exploring}
Zih-Ching Chen, Chin-Lun Fu, Chih-Ying Liu, Shang-Wen~Daniel Li, and Hung-yi Lee,
\newblock ``Exploring efficient-tuning methods in self-supervised speech models,''
\newblock in {\em 2022 IEEE Spoken Language Technology Workshop, SLT}, 2023, pp. 1120--1127.

\bibitem{huang2022improving}
Jiawen Huang, Emmanouil Benetos, and Sebastian Ewert,
\newblock ``Improving lyrics alignment through joint pitch detection,''
\newblock in {\em IEEE International Conference on Acoustics, Speech and Signal Processing, ICASSP}, 2022, pp. 451--455.

\bibitem{vaswani2017attention}
Ashish Vaswani, Noam Shazeer, Niki Parmar, Jakob Uszkoreit, Llion Jones, Aidan~N Gomez, {\L}ukasz Kaiser, and Illia Polosukhin,
\newblock ``Attention is all you need,''
\newblock {\em Advances in neural information processing systems}, vol. 30, 2017.

\bibitem{citeloss}
Rif A.~Saurous et~al,
\newblock ``The story of audioset,'' \url{http://www.cs.tut.fi/sgn/arg/dcase2017/documents/workshop_presentations/the_story_of_audioset.pdf},
\newblock 2017.

\bibitem{onsetsAndframes}
Curtis Hawthorne, Erich Elsen, Jialin Song, Adam Roberts, Ian Simon, Colin Raffel, Jesse Engel, Sageev Oore, and Douglas Eck,
\newblock ``Onsets and frames: Dual-objective piano transcription,''
\newblock in {\em Proceedings of the 19th International Society for Music Information Retrieval Conference, ISMIR}, 2018, pp. 50--57.

\bibitem{raffel2014mir}
Colin Raffel, Brian McFee, Eric~J Humphrey, Justin Salamon, Oriol Nieto, Dawen Liang, and Daniel~PW Ellis,
\newblock ``mir\_eval: A transparent implementation of common mir metrics,''
\newblock in {\em Proceedings of the 15th International Society for Music Information Retrieval Conference, ISMIR}, 2014, pp. 367--372.

\bibitem{mesaros2016metrics}
Annamaria Mesaros, Toni Heittola, and Tuomas Virtanen,
\newblock ``Metrics for polyphonic sound event detection,''
\newblock {\em Applied Sciences}, vol. 6, no. 6, pp. 162, 2016.

\bibitem{ycart2020investigating}
Adrien Ycart, Lele Liu, Emmanouil Benetos, and Marcus Pearce,
\newblock ``Investigating the perceptual validity of evaluation metrics for automatic piano music transcription,''
\newblock {\em Transactions of the International Society for Music Information Retrieval}, 2020.

\end{thebibliography}
\end{CJK*}
\end{document}